\documentclass{PoS}

\usepackage{amssymb}
\usepackage{amsmath}
\usepackage{graphicx}
\usepackage{booktabs}
\usepackage{color}
\usepackage{textcomp}
\usepackage{multirow}
\usepackage{bm}
\usepackage{caption}
\usepackage{subcaption}
\usepackage{tikz}
\usepackage{longtable}
\usepackage{colortbl}

\title{Non-Perturbative Gauge-Higgs Unification in Five Dimensions}

\ShortTitle{Non-Perturbative Gauge-Higgs Unification in Five Dimensions}

\author{\speaker{Graham Moir}, Maurizio Alberti, Francesco Knechtli \\
        Department of Physics, Bergische Universit\"{a}t Wuppertal, Gaussstr. 20, D-42119 Wuppertal, Germany \\
        E-mail: \email{gmoir@uni-wuppertal.de} }
\author{Nikos Irges \\
Department of Physics, National Technical University of Athens Zografou Campus, GR-15780 Athens, Greece}

\abstract{We study the phase diagram and mass spectrum of an $SU(2)$ Gauge-Higgs Unification scenario on a five-dimensional orbifold.
We observe spontaneous symmetry breaking within the Higgs phase of the theory and, in the vicinity of a newly discovered phase, 
we find that the ratio of Higgs to gauge boson masses takes a Standard Model-like value. Precisely in this region of the phase diagram, we observe
dimensional reduction via localisation.}

\FullConference{The 33rd International Symposium on Lattice Field Theory\\
                 14-18 July  2015\\
                 Kobe International Conference Center, Kobe, Japan}

\begin{document}

\section{Introduction}

There are many possible models that solve the `problems' associated with the Higgs sector of the Standard Model. They
typically come under the heading of either supersymmetry, technicolor or extra dimensions, and until the LHC can provide a favoured
direction, all avenues must be explored. Here we study the properties of an extra-dimensional scenario known as \textit{Gauge-Higgs Unification}.
Perturbatively, these class of models are centred around the  Hosotani mechanism \cite{Hosotani:1983vn},
where the Aharonov-Bohm phase in the extra dimension plays the role of the Higgs. This mechanism has been
used in a variety of scenarios and more recently, for example, it has been used in an $SO(11)$ Grand Unification
model \cite{Hosotani:2015hoa}. Using the arena of lattice field theory, strides have been made in exploring the non-perturbative
properties of Gauge-Higgs Unification \cite{Irges:2004gy, Irges:2006hg, Cossu:2013ora, Irges:2013rya, Alberti:2015pha, eliana}.
In particular, a five-dimensional $SU(2)$ model where the extra dimension is subject to an orbifold geometry,
has been shown to exhibit spontaneous symmetry breaking giving rise to massive gauge bosons \cite{Irges:2006hg}. 

In these proceedings, we give an overview of the results presented in \cite{Alberti:2015pha}, namely, we discuss the phase diagram
and mass spectrum of a five-dimensional pure gauge $SU(2)$ theory formulated using an orbifold geometry on an anisotropic lattice. We concentrate
on the region of the phase diagram where the lattice spacing in the usual four dimensions, $a_{4}$, is less than or equal to that along the extra dimension, $a_{5}$.

\begin{figure}[t!]
  \begin{center}
   \includegraphics[width=0.6\textwidth]{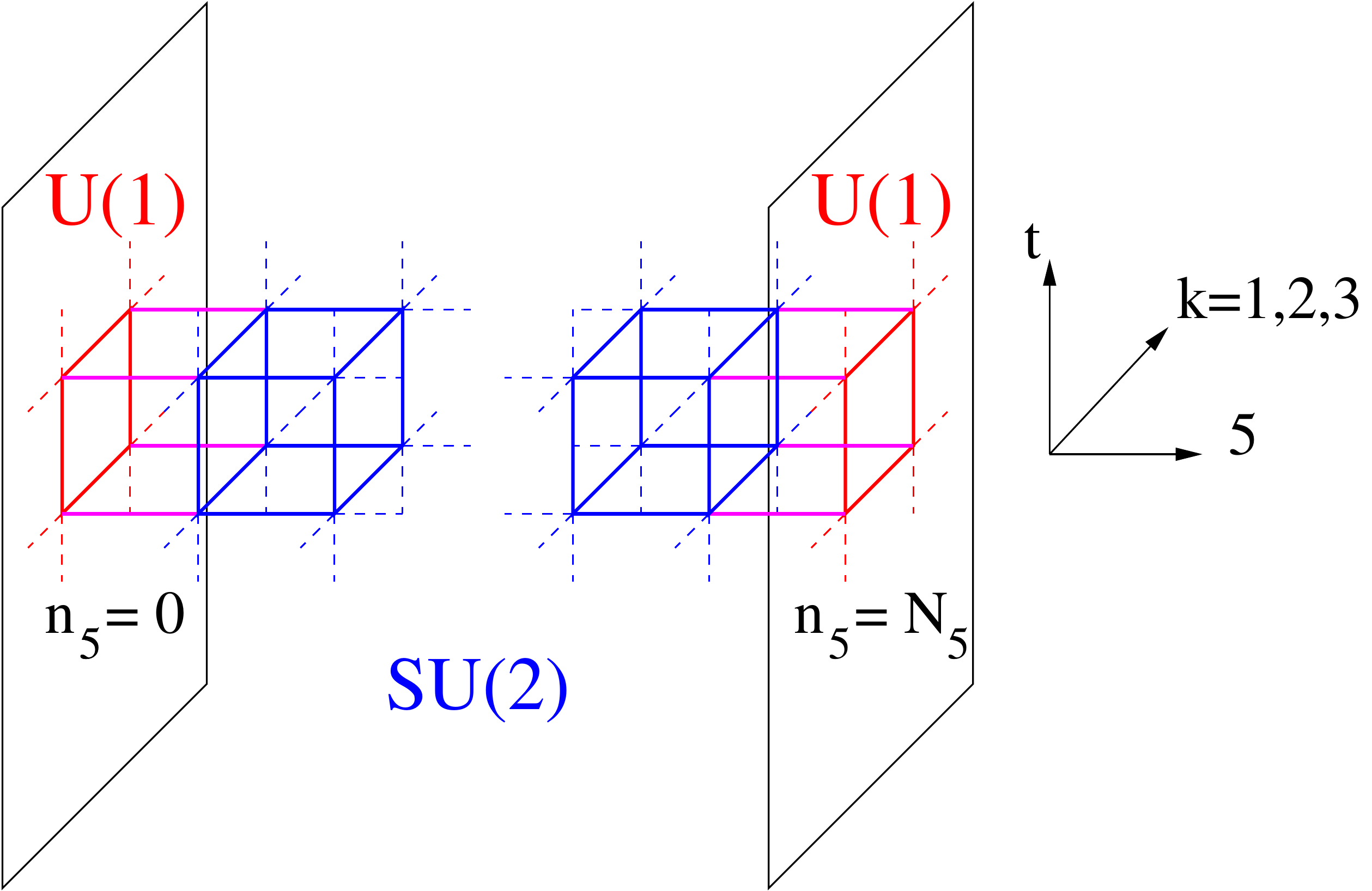}
  \end{center}
  \caption{\footnotesize A sketch of the orbifold lattice and the gauge
links: boundary $U(1)$ links are red, hybrid $SU(2)$ links
sticking to the boundaries are magenta and bulk $SU(2)$ links are blue.}
  \label{fig:latorb}
\end{figure}

The theory is defined in the domain $I = \{ n_{\mu}, 0 \leq n_{5} \leq N_{5} \}$ with volume $T \times L^{3} \times N_{5}$
corresponding to Figure \ref{fig:latorb}. The bulk $SU(2)$ gauge links are shown in blue. The gauge group is explicitly
broken from $SU(2) \rightarrow U(1)$ at the fixed points of the orbifold and the corresponding links are shown in red. The magenta links are known as `hybrid' as they
gauge transform like $SU(2)$ at one end and $U(1)$ at the other. We perform our study using a five-dimensional anisotropic Wilson action
\begin{equation}\label{eqn:wilson_action}
S_W^{orb} = \frac{\beta}{2} \sum_{n_{\mu}} {\left[ \frac{1}{\gamma}\sum_{\mu<\nu}{w~\text{tr}{\left\{1 - U_{\mu\nu}(n_{\mu}) \right\}}} + \gamma \sum_\mu{\text{tr}{\left\{ 1 - U_{\mu5}(n_{\mu}) \right\}}} \right]}~,
\end{equation}
where $w = 1/2$ for plaquettes, $U_{\mu \nu}$, living at the fixed points of the orbifold and $w = 1$ otherwise \cite{Irges:2004gy}. In the classical limit, the anisotropy parameter
$\gamma = a_{4}/a_{5}$ and $\beta = 4a_{4} / g^{2}_{5}$, where $g_{5}$ is the dimensionful continuum gauge coupling.
In what follows, we will use an equivalent pair of couplings $\beta_{4}$ and $\beta_{5}$ which are related to the couplings in equation (\ref{eqn:wilson_action}) via
\begin{equation}\label{eqn:couplings}
\beta_{4} = \frac{\beta}{\gamma} ~~~~~;~~~~~ \beta_{5} = \beta \gamma ~ .
\end{equation}
For a more complete description of the theory and lattice set-up see \cite{Alberti:2015pha}.

\section{Phase Diagram}

\begin{figure}[t!]
\begin{center}
     \includegraphics[width=0.85\textwidth]{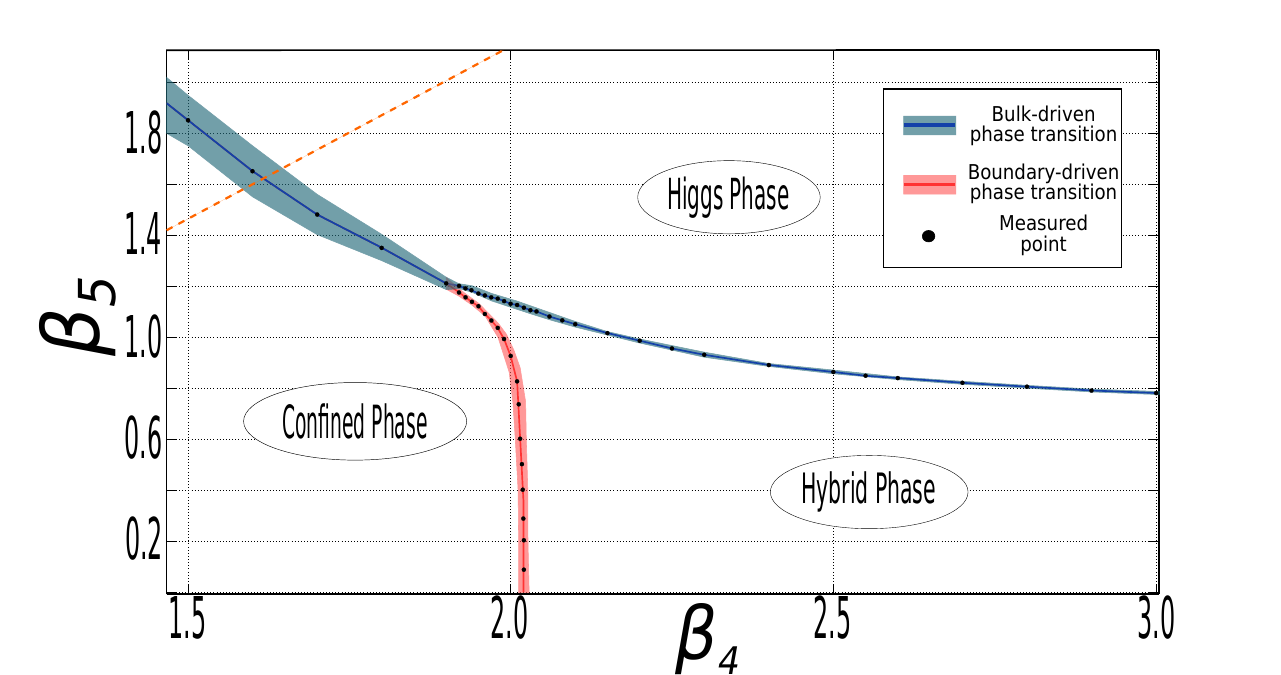}
\end{center}
    \caption{\footnotesize The phase diagram for $N_{5} = 4$ in the region of the Higgs-hybrid phase transition. The points show the location of
a first-order phase transition. The red and blue lines represent the width of the corresponding hystereses, while the dashed orange line represents $\gamma = 1$.}
  \label{fig:phase_diag}
\end{figure}

Figure \ref{fig:phase_diag} shows the phase diagram of the theory for fixed extent of the extra dimension $N_{5} = 4$, focused around the region $\gamma \leq 1$. Within our explored
parameter space, we find only first-order phase transitions which are represented by the red and blue lines. We label the blue line as \textit{bulk-driven} since it is the bulk $SU(2)$
links that drive the system into a change of phase, whereas we label the red line as \textit{boundary-driven} as it is the boundary $U(1)$ links that drive the system into a change of phase.
One important observation is that the phase structure determined here agrees on a qualitative level with that determined via mean-field \cite{Irges:2012ih,Moir:2014aha}. Another important observation
is that it is reminiscent of the $4$-D Abelian Higgs model for a Higgs of charge $q = 2$ \cite{Fradkin,Callaway}, which is the theory that the orbifold reduces to in four dimensions. 

In order to begin to label the phases within our theory, we firstly consider the expectation value of the Polyakov loop along each direction; zero expectation values imply confining dynamics
whereas non-zero values indicate deconfined dynamics. Since we find the expectation value of the Polyakov loop to be zero in all directions, we label the lower left phase of
Figure \ref{fig:phase_diag} as \textit{confined}. We label the upper region as the \textit{Higgs phase} since it exhibits deconfinement in all directions and gives a non-zero mass of the
Higgs and gauge boson. The third phase we label as \textit{hybrid} since it exhibits deconfined dynamics on the four-dimensional boundary hyperplanes and confining dynamics elsewhere; this behaviour is strongly
reminiscent of the \textit{layered phase} described in \cite{Fu}. In the rest of these proceedings, we will focus on the Higgs phase since it is the best candidate to reproduce a Standard Model-like
Higgs sector. For further discussion on the properties and features of the other phases, see \cite{Alberti:2015pha}. 

\section{The Higgs Phase}

We are interested in the properties of the Higgs phase since it is the region of the phase diagram where we can determine a Higgs mass and observe spontaneous symmetry breaking 
giving rise to a massive vector boson. We label this vector boson as the $Z$ due to its similarity to with the Standard Model $Z$ boson. The spontaneous symmetry breaking is governed
by the so-called \textit{stick symmetry} \cite{Ishiyama}, and we refer the reader to \cite{Alberti:2015pha} for a detailed explanation.
In order to determine if these types of theories can be candidate solutions to the problems surrounding the Higgs sector of the Standard Model, we determine the low-lying mass spectrum
within the Higgs ($J^{PC} = 0^{++}$) and Z boson ($J^{PC} = 1^{--}$) channels. The Standard Model has a ratio of Higgs to $Z$ boson masses $\rho \approx 1.38$ and in order for Gauge-Higgs Unification
scenarios to be phenomenologically viable, they should achieve a similar value for a range of physically similar phase space parameters; an indication that this is achievable in our current $SU(2)$ model
would encourage further explorations into larger models that can account for all the degrees of freedom of the Standard Model. For example, an $SU(3)$ theory could reproduce the full Higgs sector as
it would break to $SU(2) \times U(1)$ at the fixed points of the orbifold.

\subsection{Spectrum}

\begin{figure}[t!]
  \begin{center}
  \includegraphics[width=0.75\textwidth]{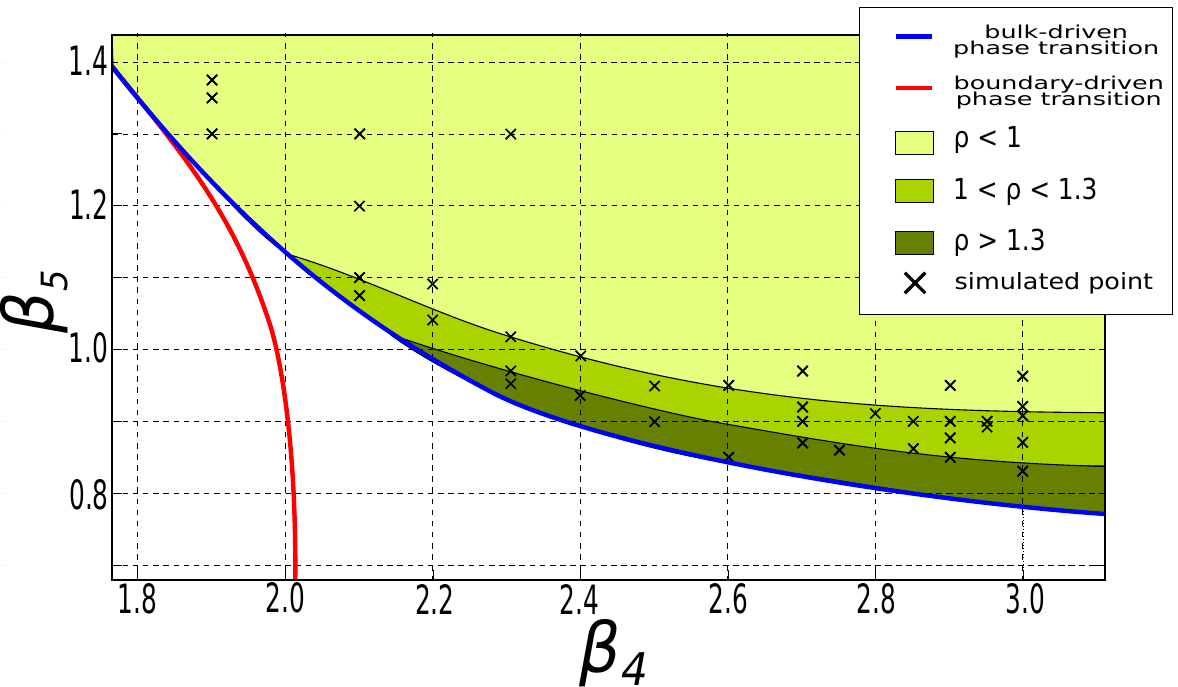}
  \end{center}
  \caption{\footnotesize Summary of our spectroscopic calculations for $\gamma < 1$ and $N_{5} = 4$ close to the Higgs-hybrid phase transition. The black points indicate
                         the location of a spectroscopic calculation within the phase diagram. The green shading indicates the $\rho \equiv m_{H}/m_{Z}$ value obtained for
                         a given calculation. The lightest shade corresponds to $\rho < 1$, the middle shade to $1 \leq \rho \leq 1.3$ and the darkest shade
                         corresponds to $\rho > 1.3$.}
  \label{fig:phase_diagram_with_rho}
\end{figure}

We determine the low-lying spectrum in both the Higgs and $Z$ boson channels by solving a generalised eigenvalue problem with a large basis of interpolating operators
as described in \cite{Alberti:2015pha}. When the gauge couplings in the four-dimensional hyperplanes and along the extra dimension are equal (i.e. along the dashed orange line in 
Figure \ref{fig:phase_diag}), we find that the mass of the $Z$ boson is always heavier than that of the Higgs. However, as we lower the inverse coupling along the extra dimension, $\beta_{5}$,
we find a tendency of this hierarchy to become more Standard Model-like.

Figure \ref{fig:phase_diagram_with_rho} shows a summary of our spectroscopic calculations for fixed $N_{5} = 4$ within the vicinity of the Higgs-hybrid phase transition (blue line). We observe that
once we keep the coupling along the four-dimensional hyperplanes $\beta_{4} > 2.02$ and are close enough to the phase transition, the Higgs mass is heavier than that of the $Z$. The value of
$\beta_{4} = 2.02$ is no coincidence as it is the value within this system for which the boundary $U(1)$ links \textit{naturally} deconfine; below that value a decoupled four-dimensional $U(1)$
hyperplane would exhibit confined dynamics and above it exhibit deconfined dynamics. Furthermore, we find that the theory achieves a Standard Model-like ratio $\rho \equiv \sim 1.38$ of Higgs to $Z$ boson
masses for a wide range of parameters, highlighted by the darkest green shading in Figure \ref{fig:phase_diag}. \\

\begin{figure}[t!]
  \begin{center}
   \includegraphics[width=0.85\textwidth]{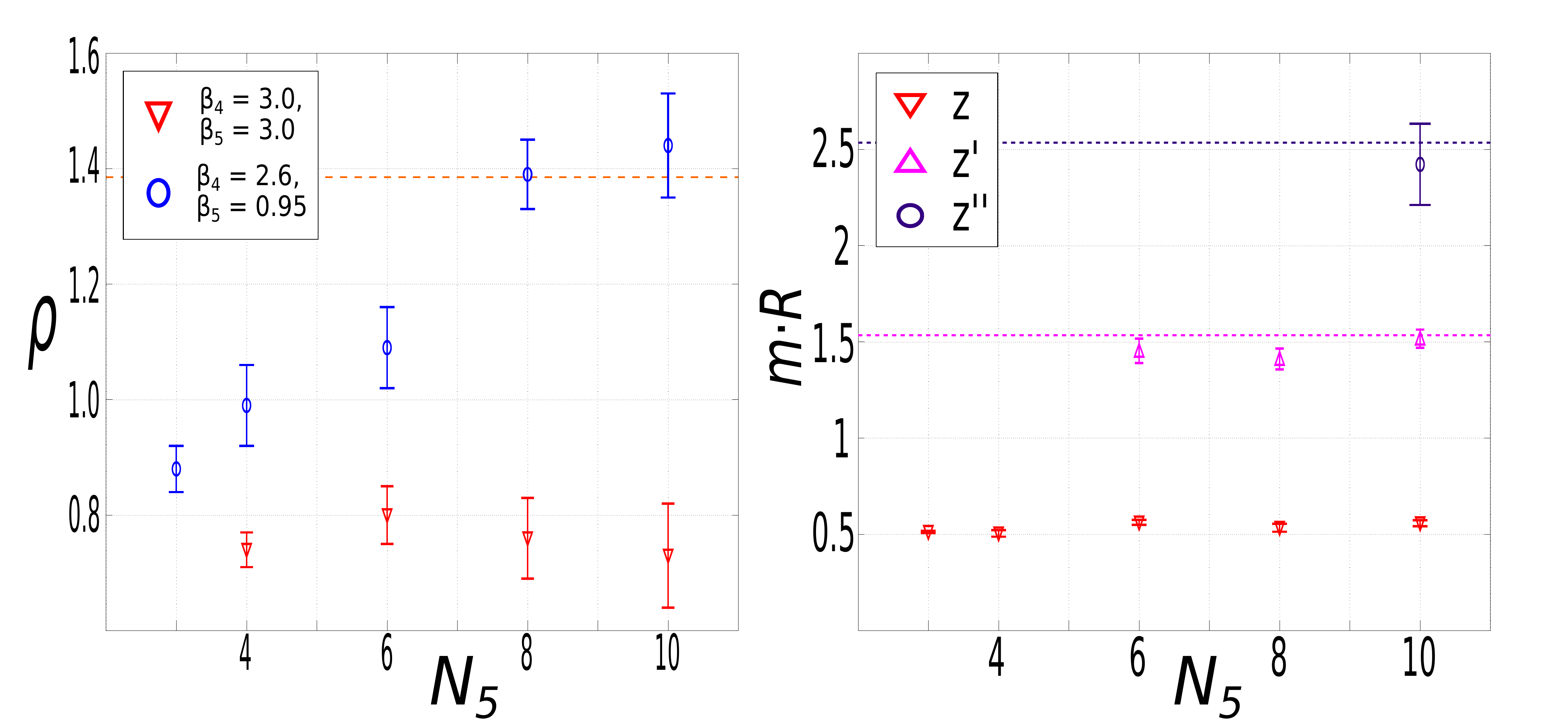}
  \end{center}
  \caption{\footnotesize The left panel shows the ratio of Higgs to $Z$ boson masses $\rho$ for various values of $N_{5}$ for two sets of fixed inverse gauge couplings
  $\{\beta_{4},\beta_{5}\} = \{2.6, 0.95\} , \{3.0, 3.0\}$. The right panel shows the $N_{5}$ dependence of the $Z$ channel in units of the radius of the extra dimension for the same set of couplings.
   The orange dashing in the left panel is the Standard Model value of $\rho$, whereas the dashed lines in the right panel show where Kaulza-Klein excitations would appear.}
  \label{fig:N5_scan}
\end{figure}

The left panel of Figure \ref{fig:N5_scan} shows the dependence of the ratio $\rho$ on the extent of the extra dimension, $N_{5}$, at fixed $\beta_{4} = 2.6$ and $\beta_{5} = 0.95$. At $N_{5} = 4$, this point
lies within the $\rho < 1$ region of Figure \ref{fig:phase_diagram_with_rho} and has a value $\rho \approx 0.85$. However, as $N_{5}$ is increased it is clear that $\rho$ takes on a Standard Model-like
value, which is given by the orange dashing in the figure. Since we find that the location of the phase transition becomes $N_{5}$ independent for $N_{5} > 4$, the region of Standard Model-like parameter space
increases with $N_{5}$. This will be an important feature when constructing lines of constant physics; since we find only first order phase transitions, we expect that this theory can only be treated as an effective one and hence, we must find a region of parameter space that scales appropriately with the cut-off while keeping the physics constant.

The right panel of Figure \ref{fig:N5_scan} shows the $N_{5}$ dependence, in units of the radius of the extra dimension, of the $Z$ boson mass and its excitations $Z'$ and $Z''$ (where we could robustly determine them). It is clear that we see a $1/N_{5}$ behaviour of each particle and that the excitations behave like a Kaluza-Klein tower, that is, that excitations $m^{KK}_{i+1}$ appear at an energy scale
$1/R$ higher than $m^{KK}_{i}$.

\subsection{Dimensional Reduction}

Extra-dimensional models are only viable if the extra dimensions are hidden at energy scales of the Standard Model. We are therefore only interested in regions of parameter
space that dimensionally reduce to a four-dimensional Standard Model-like Higgs sector. By determining the potential between a pair of static charges of the respective gauge group
within layers orthogonal to the fifth dimension, not only can we determine the dimensionality of the system but we can also determine the dynamics of the system. As described in
\cite{Alberti:2015pha}, we fit different potential shapes to our extracted potential. Within the Higgs phase, these shapes include $4$-D Yukawa, $5$-D Yukawa, $4$-D Coulomb and $5$-D Coulomb.

\begin{figure}[t!]
  \begin{center}
   \includegraphics[width=0.85\textwidth]{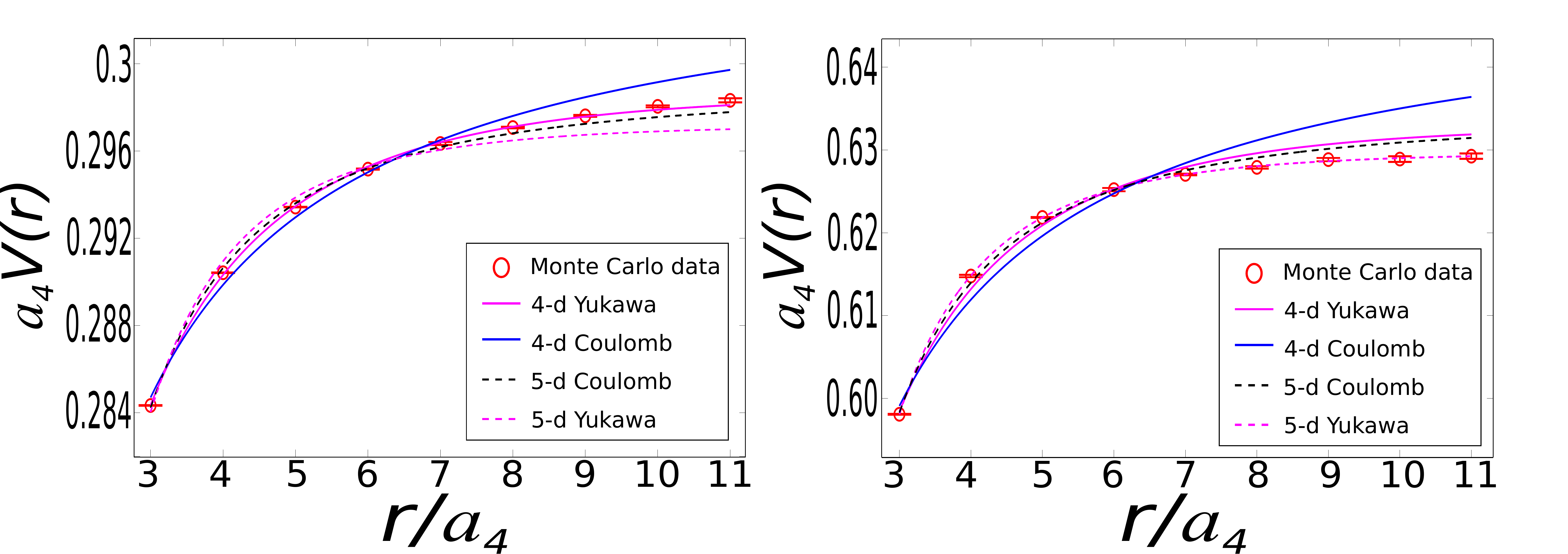}
  \end{center}
  \caption{\footnotesize The left (right) panel shows fits to the potential determined on the $n_{5} = 0$ boundary ($n_{5} = 2$ bulk) hyperplane orthogonal to the extra dimension for $\{\beta_{4},\beta_{5}\} = \{2.1, 1.075\}$.}
  \label{fig:stat_pot}
\end{figure}

For isotropic couplings, where we find that the mass of the $Z$ is always heavier than that of the Higgs, we find no evidence of dimensional reduction and find that the potential determined on
each hyperplane can only be described by a $5$-D Yukawa form. However, in the region where we find a Standard Model-like spectrum we find dimensional reduction via
\textit{localisation} on the $U(1)$ boundary hyperplanes of the orbifold. The left panel of Figure \ref{fig:stat_pot} shows fits of different potential types to our extracted potential on the
four-dimensional boundary ($n_{5} = 0$) hyperplane for $\beta_{4} = 2.1$ and $\beta_{5} = 1.075$. It is clear that the favoured fit is of $4$-D Yukawa type; the other fits have $\chi^{2}$ per degree of
freedom values at least an order of magnitude higher than that of the $4$-D Yukawa. In the right panel we show fits to the extracted potential on the middle ($n_{5} = 2$) four-dimensional hyperplane of
the orbifold. Here we see that only a $5$-D Yukawa fit is possible and again the other fits have $\chi^{2}$/d.o.f values at least an order of magnitude higher. This behaviour is identical on all bulk
hyperplanes. 

One consistency check that can be preformed is to measure the mass of the $Z$ boson obtained from the fit to the $4$-D Yukawa potential and compare it to the value obtained independently in our spectroscopic calculation; a differing value would suggest that the fit is not optimal while a matching value confirms that the fit to the potential is correctly described. The value in lattice units obtained from the
spectroscopic calculation is $a_{4}m_{z} = 0.268(3)$ and the value obtained from the $4$-D Yukawa fit is $a_{4}m_{z} = 0.26$. This remarkable agreement confirms the dimensionality of the boundary 
to be four, implying dimensional reduction of the system via localisation.

\section{Conclusions and Outlook}

We have demonstrated that a five-dimensional $SU(2)$ Gauge-Higgs Unification scenario embedded within an orbifold geometry possesses a large region of parameter space that dimensionally
reduces via localisation and exhibits a Standard Model-like ratio of Higgs to $Z$ boson masses. We also find a $Z'$ particle present in the spectrum at the scale of the extra dimension. Our
next step is to construct lines of constant physics within the Standard Model-like parameter space. Work in this direction has already begun. 

Encouraged by the results of this study, we plan an exploration into larger models that contain all the degrees of freedom necessary for the Standard Model Higgs sector, namely an $SU(3)$ theory which
breaks to $SU(2) \times U(1)$ on the boundary.

\section*{Acknowledgements}

This work was funded by the Deutsche Forschungsgemeinschaft (DFG) under contract KN 947/1-2. In particular G. M. acknowledges full support from the DFG. The Monte Carlo simulations were carried out on Cheops, one of the supercomputers funded by the DFG at the RRZK computing centre of the University of Cologne and on clusters at the University of Wuppertal. We thank both Universities.

\end{document}